# Two-photon interference and coherent control of single InAs quantum dot emissions in an Ag-embedded structure

X. Liu,[1,2,a] H. Kumano,[1] H. Nakajima,[1] S. Odashima,[1] T. Asano,[1] T. Kuroda,[2] and I. Suemune[1]

[1]*Research Institute for Electronic Science, Hokkaido University, Sapporo 001-0021, Japan*

[2]*National Institute for Materials Science, 1-1 Namiki, Tsukuba 305-0044, Japan*

**Abstract**- We have recently reported the successful fabrication of bright single-photon sources based on Ag-embedded nanocone structures that incorporate InAs quantum dots. The source had a photon collection efficiency as high as 24.6%. Here we show the results of various types of photonic characterizations of the Ag-embedded nanocone structures that confirm their versatility as regards a broad range of quantum optical applications. We measure the first-order autocorrelation function to evaluate the coherence time of emitted photons, and the second-order correlation function, which reveals the strong suppression of multiple photon generation. The high indistinguishability of emitted photons is shown by the Hong-Ou-Mandel-type two-photon interference. With quasi-resonant excitation, coherent population flopping is demonstrated through Rabi oscillations. Extremely high single-photon purity with a $g^{(2)}(0)$ value of 0.008 is achieved with $\pi$-pulse quasi-resonant excitation.

---

[a] Electronic mail: LIU.Xiangming@nims.go.jp

# 1. Introduction

Single-photon sources are of great interest and importance in terms of realizing absolutely secure quantum key distribution (QKD) and long-distance quantum communications over the widespread faint laser sources whose intensity is attenuated to a single-photon level.[1] This is because a faint laser source has a Poisson distribution of photons, which makes QKD insecure. In addition, the strong attenuation of laser light leads to a significant reduction in the key generation rate. These problems are avoided by use of single-photon sources, which employ atom-like two-level systems with a ground state and an excited state. Among the two-level single-photon sources,[2-4] epitaxially grown semiconductor quantum dots (QDs) have been regarded as promising candidates because of their electronic-drive capability and the tunability of their emission wavelengths. Furthermore, QDs have also been used for the generation of entangled photon pairs via the cascade emission of two photons following the recombination of a biexciton and the associated exciton.[5]

One of the major issues for a single-photon source based on single QDs is to increase the photon collection efficiency, which is defined as the fraction of photons collected by the first lens of the optical setup. Generally, the collection efficiency for a single QD embedded in an unprocessed semiconductor sample is relatively small (< 1%).[6] The large refractive index difference between the QD matrix and the surrounding air severely limits photon collection due to the total internal reflection at the interface. Important progress has been made towards the development of bright QD-based single-photon sources. The tailored structures include distributed Bragg reflector (DBR) microcavities with pillar structures[7] or lateral trenches for optical confinement,[8] photonic nanowires,[9,10] horn structures,[11] and circular dielectric gratings.[12] The figures of merit of these photon sources are listed in Table I. Other structures have also been proposed including micro-pyramids,[13] photonic crystal waveguides,[14] and trumpet structures.[15] Although some tailored structures provide relatively high collection efficiencies, complicated structural processes are usually required, making these structures unstable and inappropriate for the optical fiber contacts utilized in fiber-based networks. Recently, to achieve a bright single-photon source with good structural robustness for the fiber contact,[16] we have proposed metal-coated semiconductor nanostructures (nanopillars[17-19] and nanocones[20]) incorporating InAs QDs that utilize metals with high optical reflectance. We embed the semiconductor nanostructures completely into the metals, which makes the nanostructures mechanically stable. The flat surfaces enable us to extract photons efficiently from the source to the outer optics. Thus, we can expect to achieve a fiber-based photon source with long-term stability through the direct contact of this kind of photon source to a single-mode fiber. In particular, we have shown that the inclination of a pillar sidewall embedded in Ag (Ag-embedded nanocone structure) facilitates the improvement of photon collection efficiency and we demonstrated experimentally a photon collection efficiency of 24.6 % for a lens with a numerical aperture (NA) of 0.42.[20]

TABLE I. Figures of merit of various photon sources

| Sources | NA of lens | $g^{(2)}(0)$ at saturation | Collection efficiency (%) | Indistinguishability (visibility) |
|---|---|---|---|---|
| DBR microcavities[7] | 0.5 | >0.3 | 8 | … |
| Lateral trenches[8] | 0.4 | 0.4 | 38 | … |
| Photonic nanowires[9] | 0.4 | <0.008 | 35 | … |
|  | 0.75 |  | 72 | … |
| Tailored nanowires[10] | 0.75 | 0.36 | 42 | … |
| Horn structure[11] | 0.55 | 0.19 | 10.9 | … |
| Circular dielectric grating[12] | 0.42 | … | 10 | … |
| Nanopillar[19] | 0.42 | 0.31 | 18 | … |
| Nanocone[20] | 0.42 | 0.3 | 24.6 | … |
| Light-emitting diode[21] | … | … | … | 0.33 |
| Micropillar cavity[22] | … | … | … | 0.90 |
| Microdisk cavity[23] | … | … | … | 0.33 |

In this paper, we present various photonic characterizations of our Ag-embedded nanocone structures incorporating InAs QDs, which were originally developed as on-demand single-photon sources. First, we confirm the single-photon nature of the QD photon source by measuring the anti-bunching behavior in the second-order correlation function. We show that the improved photon collection efficiency of up to 19% is reproducible in this metal-semiconductor structure. Second, we determine the coherence time of emitted single photons and discuss the dominant dephasing processes responsible for its limitation. Third, we demonstrate the high indistinguishability of emitted photons by using Hong-Ou-Mandel (HOM) two-photon interference.[21-23] The indistinguishability of photons transmitted in a quantum channel is a fundamental requirement when performing a Bell-state measurement, which makes it possible to construct quantum repeaters for long-distance quantum communications.[1] We prove that our photon source can combine high brightness and indistinguishability simultaneously. Finally, we present a result for the coherent control of quantum states by observing exciton Rabi oscillation. In contrast to the incoherent excitation approach that is conventionally used for the excitation of QD sources, the present coherent excitation approach enables us to prepare the excited state deterministically with near-unity probability. We obtain an extremely high purity ($g^{(2)}(0)=0.008$) for single photons under $\pi$-pulse excitation. These characterizations prove that our Ag-embedded nanocone structure incorporating single InAs QDs can serve as a versatile high-performance source of single photons.

## 2. Sample and measurement setup

The single-photon source studied here is based on a single InAs QD grown on a GaAs (100) substrate. A tailored GaAs nanocone prepared with electron-beam lithography (EBL) and reactive ion etching (RIE) processes was deposited with Ag, followed by mechanical polishing and sequential etching to remove the substrate. The QD growth and the processes used to fabricate the metal-embedded nanostructure are described in detail in Ref. 20. Figure 1 is a schematic of the measurement setup. The sample was mounted inside a cryostat and cooled down to 20 K. A Ti: Sapphire laser under CW or mode-locked operation with a repetition rate of 76 MHz and a pulse duration of ~5 ps was focused on one of the nanostructures through an objective lens with NA=0.42. Photons spontaneously emitted from QDs were collected with the same objective lens and then sent to a 64-cm-long triple monochromator equipped with a liquid-nitrogen-cooled Si charge-coupled-device (CCD) detector. A tunable band-pass filter (BPF) with a full width at half maximum (FWHM) of 0.5 nm was used to select a single emission line. An additional long-pass filter (LPF) and short-pass filter (SPF) were used to reduce unwanted photons during autocorrelation measurements with a Hanbury-Brown-Twiss (HBT) setup[24] or HOM measurements. A Michelson interferometer was employed for the first-order coherence measurements.

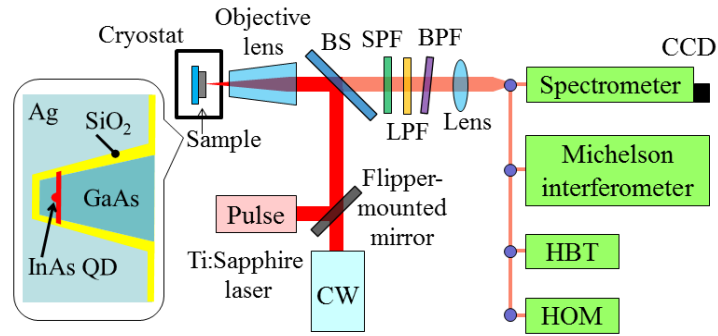

FIG. 1. Schematic of the measurement setup.

## 3. Single QD spectroscopy and photon anti-bunching

Figure 2(a) shows the photoluminescence (PL) spectrum measured from a nanocone structure incorporating an InAs QD with 844-nm CW excitation. The dominant peak at 935.1 nm with a linewidth of 113 µeV showed no fine-structure splitting (FSS) and was attributed to a negatively charged exciton ($X^-$).[25] The weaker emission at 931.0 nm was attributed to a neutral exciton emission ($X^0$) as its FSS was 66 µeV and its intensity increased linearly with the excitation power (not shown).[25] The linewidth of the $X^0$ emission was 191 µeV.

To measure the exciton lifetime ($\tau_r$) of the emission $X^-$, the single QD was excited with an 844-nm pulse. The $X^-$ emission photons, which were spectrally selected with the tunable BPF, were coupled into a single-mode fiber and then directed into one superconducting single-photon detector (SSPD). The transient decay was measured by using cross-correlation measurements with the laser pulse directed into another SSPD. The measured time decay of the $X^-$ line is shown in Fig. 2(b). The inset is the instrument response function (IRF) measured with the HBT setup shown in Fig. 2(c), and the IRF response time deduced from the fitted FWHM was 146 ps. The solid line in Fig. 2(b) is the convolution of a single exponential function with the IRF, and the lifetime of the $X^-$ emission was 1.24 ns. This value is comparable to the previously reported exciton lifetime of optically excited InAs/GaAs QDs.[18,26]

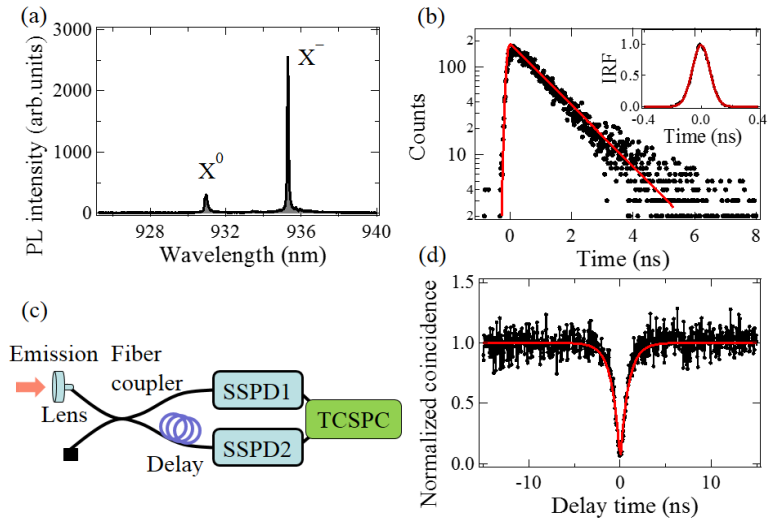

FIG. 2. (a) PL spectrum measured from a single QD at 20K. The emission lines $X^0$ and $X^-$ correspond to the neutral exciton and the negatively charged exciton, respectively. (b) Time-resolved PL decay of the $X^-$ line. The solid curve is the convolution of single exponential decay with IRF. The IRF shown in the inset is fitted with Gaussian function. (c) HBT setup based on optical fibers for measuring second-order correlation function. An optical delay of 25 ns was inserted. TCSPC represents the time-correlated single-photon counting. (d) Normalized coincidence of the $X^-$ line as a function of the delay time. The solid curve is the fit to the experimental data.

The single-photon nature of the QD emitter was investigated by measuring the second-order correlation function: $g^{(2)}(\tau)=\langle \hat{E}(t)\hat{E}(t+\tau)\hat{E}^+(t+\tau)\hat{E}^+(t)\rangle/\langle \hat{E}(t)\hat{E}^+(t)\rangle^2$, where $\hat{E}(t)$ and $\hat{E}^+(t)$ are the electric field operators and $\tau$ is the time delay. In the photon correlation measurements under CW excitation, the spectrally selected X$^-$ photons were divided by a fiber coupler as shown in Fig. 2(c). An optical delay of 25 ns was inserted into one path to shift the zero time delay. $g^{(2)}(\tau)$ under CW excitation can be expressed as[27]

$$g^{(2)}(\tau) = 1-[1-g^{(2)}(0)]\exp(-|\tau|/\tau_{tot}), \qquad (1)$$

where $\tau_{tot}$ is the time constant given by $1/\tau_{tot}=1/\tau_r+1/\tau_p$ and $\tau_p$ is the inverse of the pumping rate. $g^{(2)}(0)$ represents the background photon emission and multi-photon emission from the QD, and $g^{(2)}(0)=0$ indicates a perfect single-photon operation. The normalized coincidence of the X$^-$ emission measured against the delay time under CW excitation is shown in Fig. 2(d). A strong suppression of the uncorrelated background emission was observed with a value of $g^{(2)}(0)=0.05$. This value is much smaller than 0.5, which is the criterion of the quantum nature, and therefore indicates that the X$^-$ emission generated from the single QD exhibited a single-photon nature. A lower $g^{(2)}(0)$ is expected with quasi-resonant excitation as discussed later.

We also measured the collection efficiency of photons emitted from the nanocone embedded in Ag. The method is the same as that described in Ref. 20. The nanocone was excited with an 800-nm pulsed laser at a repetition frequency of 76 MHz. The average excitation power of 4 μW almost saturated the X$^-$ emission at a detection rate of 180.9 KHz and the $g^{(2)}(0)$ value was estimated to be 0.22. By multiplying $\sqrt{1-g^{(2)}(0)}$ to correct the multi-photon emission probability and dividing by the photon detection efficiency of 0.011 for the measurement setup, the single-photon flux collected at the objective lens was 14.5 MHz. This corresponds to a photon collection efficiency of 19%, which almost reproduces our previous result (24.6%).[20] The slight decrease in photo collection efficiency was probably due to the imperfection in the sidewall introduced during the fabrication process. We can expect a further increase in the collection efficiency by replacing the current objective with one that has a larger numerical aperture.[16]

## 4. Coherence property of single photons

The coherence property of the generated single photons was investigated by measuring the first-order autocorrelation function $g^{(1)}(\tau)$,[28-30] which can be expressed as $g^{(1)}(\tau)=\langle \hat{E}(t)\hat{E}^+(t+\tau)\rangle/\langle \hat{E}(t)\hat{E}^+(t)\rangle$. Using the Michelson interferometer shown in Fig. 3(a), we recorded the interference fringes by rotating the phase shifter to vary one of the arm lengths over several wavelengths at each path length difference set by the translation stage. The visibility of the interference fringes is dependent on the time delay $\tau$ of the two optical paths and can be given by[31]

$$V^{(1)}(\tau) = \frac{I_{max} - I_{min}}{I_{max} + I_{min}} = |g^{(1)}(\tau)|, \qquad (2)$$

where $I_{max}$ and $I_{min}$ are the interference maximum and minimum, respectively. An example of the interference fringes measured at a position near zero time delay is given in the inset of Fig. 3(b), which provides a fringe visibility as high as 0.84. The deviation of the visibility from the ideal value of 1 is mainly due to the cross-polarization coupling inside the fiber and the resultant elliptical polarization at the output. As shown in Fig. 3(b), the change in the magnitude of the visibility follows the exponential decay $V^{(1)}(\tau) = V^{(1)}(0)exp(-|\tau|/\tau_c)$, where $\tau_c$ is the coherence time. The fitting shows that the coherence time is 49 ps for the X⁻ emission. The measured coherence time of $X^0$ was 35 ps (not shown). This shorter coherence time was attributed to the impact of spectral diffusion on the line broadening.[32]

In an ideal QD emitter, the coherence time should be twice the recombination lifetime. However, the coherence time is shortened as a result of the exciton dephasing including the exciton–phonon interaction and exciton–carrier scattering. At cryogenic temperatures, the exciton-phonon interaction has only a small effect on dephasing, while the exciton-carrier scattering dominates the dephasing. Random carrier motion in the vicinity of the QD induces spectral diffusion, which leads to an additional reduction of the coherence time. The spectral diffusion is probably enhanced in the nanocone structures, which have a large surface/volume ratio compared with a bulk system. Improvements in fabrication techniques are expected to increase the coherence time of emitted photons.

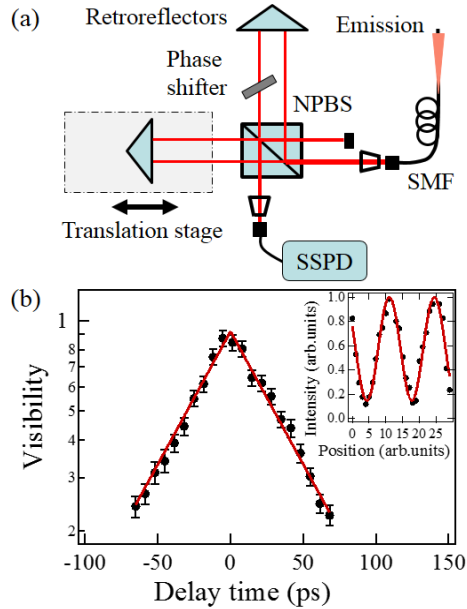

FIG. 3. (a) Experimental setup of Michelson interferometer for measuring the coherence time. NPBS: non-polarizing beam splitter. The phase shifter was introduced to record the interference fringes at a fixed optical-path length difference set by the translation stage. (b) The visibility as a function of delay time for the X⁻ emission. The inset shows an example of interference fringes at a position close to zero time delay.

## 5. Indistinguishability of single photons

To prove the indistinguishability of two consecutively emitted photons, we carried out an HOM two-photon interference experiment[21] employing a Mach-Zehnder interferometer under CW excitation as shown in Fig. 4(a). A polarizer was inserted before the first non-polarizing beam splitter (NPBS1) to generate vertically polarized photons. The interferometer had a fixed time delay of $\Delta t = 5.8$ ns between the two optical paths. The short path included a half-wave plate for rotating the polarization. For the configuration of the vertical and horizontal orthogonal polarizations in the two paths, the photons are distinguishable and the second-order correlation function at zero time delay, $g_\perp^{(2)}(0)$, is equal to 0.5 for a perfectly balanced interferometer. With a parallel (identical) polarization configuration, the two photons are indistinguishable and bunch at one of the output ports. For a pure single-photon emitter with perfect indistinguishability, the central dip ($g_\parallel^{(2)}(0)$) should be 0. The formulae for $g_\parallel^{(2)}(\tau)$ and $g_\perp^{(2)}(\tau)$ are as follows:[21]

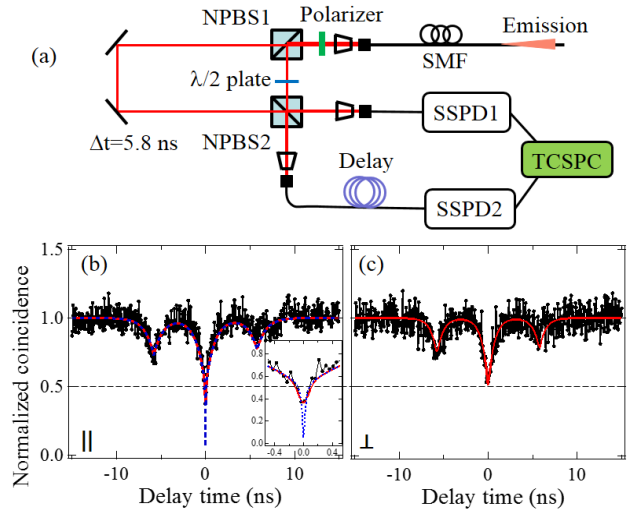

FIG. 4. (a) Schematic of CW two-photon interference measurement. A fixed time delay between two optical paths was 5.8 ns. (b) and (c) are the normalized two-photon interference coincidences of the X⁻ line under parallel and orthogonal polarizations, respectively. The inset in (b) is the zoom-in near the central dip. The horizontal dashed line corresponds to the $g^{(2)}(0) = 0.5$ level. The solid curves in (b) and (c) are calculated for parallel and orthogonal polarizations, respectively, with Eqs. (3) and (4) by taking the convolution with the IRF. The dashed curve in (b) is the calculation without the convolution.

$$g_\parallel^{(2)}(\tau) = 4(T_1^2 + R_1^2)R_2T_2 g^{(2)}(\tau) + 4R_1T_1[T_2^2 g^{(2)}(\tau - \Delta t) + R_2^2 g^{(2)}(\tau + \Delta t)] \times (1 - Ve^{-2|\tau|/\tau_c}), \quad (3)$$

$$g_\perp^{(2)}(\tau) = 4(T_1^2 + R_1^2)R_2T_2 g^{(2)}(\tau) + 4R_1T_1[T_2^2 g^{(2)}(\tau - \Delta t) + R_2^2 g^{(2)}(\tau + \Delta t)]. \quad (4)$$

$g^{(2)}(\tau)$ is given in Eq. (1). $R_i$ and $T_i$ ($i = 1, 2$) are the reflection and transmission intensity coefficients of NPBS1 and NPBS2. $V$ is a function that is dependent on the spatial overlap of the photon wave packets at NPBS2. Figure 4(b) and (c) show the CW two-photon interference $g_\parallel^{(2)}(\tau)$ and $g_\perp^{(2)}(\tau)$ of the emission line X$^-$ for parallel and orthogonal polarizations, respectively. The values at zero time delay are $g_\parallel^{(2)}(0) = 0.37$ and $g_\perp^{(2)}(0) = 0.53$. The inequality of the two dips at ±5.8 ns indicates that NPBS2 was not perfectly balanced. The solid lines in Figs. 4(b) and (c) were calculated with Eqs. (3) and (4) by taking the IRF into account. The optical properties of the NPBSs were measured separately: $R_1 = 0.46$, $T_1 = 0.54$, $R_2 = 0.48$, $T_2 = 0.52$. For the calculation, we used the coherence time of $\tau_c = 49$ ps obtained from the first-order coherence measurements and assumed that the two-photon overlap $V$ was 100%.[21] It can be seen that the fitting curves are in good agreement with the measured data. The maximum two-photon interference visibility $V_{HOM} = [g_\perp^{(2)}(0) - g_\parallel^{(2)}(0)] / g_\perp^{(2)}(0) = 0.30$ was obtained. The dashed line in Fig. 4(b), calculated without the influence of the IRF, indicates that the value of $g_\parallel^{(2)}(0) = 0.37$ is limited by the IRF. With an improvement in IRF time resolution, $g_\parallel^{(2)}(0)$ will approach a value of 0.06. Then the two-photon interference visibility will be asymptotic to the value of $V_{HOM} = 0.89$. Similar high visibilities derived from the de-convolution of the IRF function are given in Ref. 33. It should be noted that the assumption of a $V = 100\%$ two-photon overlap has been generally accepted in the CW two-photon interference measurements[21-23] but this leads to an overestimation of the $V_{HOM}$ values.

## 6. Coherent control of single-photon emission

Generally, the purity of a single-photon emission can be improved by suppressing the recapture processes and multi-photon generation probability under quasi-resonant excitation.[19,34] Information about the absorption spectra related to the luminescent states in a QD can be obtained by measuring the PL excitation (PLE) spectra. For the QD used above, the PLE spectra measured by detecting the X$^0$ and X$^-$ emission lines are shown in Fig. 5. The measurements were performed by scanning the photon energy of the tunable excitation laser and detecting the exciton emission intensity. The PLE spectrum of X$^0$ at a wavelength of 931.0 nm shows that the first excited-state absorption at 913.1 nm dominates. We also find that absorption corresponding to 1LO- and 2LO-phonon-assisted excitation appears at wavelengths of 908 nm and 897 nm, respectively. It is interesting to note that the PL intensity of the $X^-$ emission is much weaker than that of $X^0$ especially with the excitation wavelength close to the emission lines. This demonstrates that there are rarely residual charges in the QD and the appearance of the $X^-$ emission line is due to the high probability of electron transfer from the excited surrounding area.[35,36]

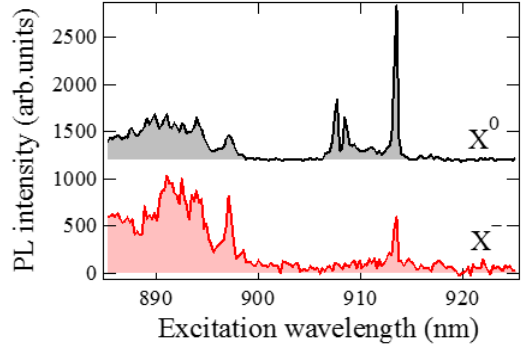

FIG. 5. PLE spectra measured on $X^0$ and $X^-$ emissions. It should be noted that the PLE spectrum of the $X^-$ line is multiplied by 5.

Figure 6(a) shows the PL spectrum of the same QD under the quasi-resonant excitation of the first excited state at 913.1 nm with a low excitation power of 3.5 μW. It can be seen that the dominant emission peak has shifted from the $X^-$ at 935.1 nm in Fig. 2(a) under off-resonant excitation to the $X^0$ at 931.0 nm. The inset in Fig. 6(a) shows the normalized photon counts of the $X^0$ emission as a function of the square root of the excitation power density with pulsed excitation at 913.1 nm. The periodic variation in intensity is known as the exciton Rabi oscillation between the exciton ground state and the QD first excited state. Similar behaviors have been observed for quasi-resonant and resonant excitation.[37,38] The PL intensity, which is proportional to the population of the excited state, undergoes sinusoidal oscillations given as $I(t) \propto \sin^2(\Theta(t)/2)$. $\Theta(t) = (\mu/h)\int_{-\infty}^{t} E(t')dt'$ is the input pulse area, where $\mu$ and $E(t')$ are the transition dipole moment and the electric-field envelope, respectively. When $t$ goes to infinity, $\Theta(\infty)$ is proportional to the square root of the excitation laser intensity. As shown in the figure, the first maximum corresponds to $\Theta = \pi$ ($\pi$-pulse excitation) at 25 $\sqrt{W/cm^2}$, which indicates that the two-level system is entirely in the excited state if the dephasing or relaxation is negligible. The first minimum corresponds to $\Theta = 2\pi$, which means that the system has been entirely back to the ground state. At any other excitation intensity, the system is in the superposition state. From the experimental data, we obtain a transition dipole moment of $\mu = 11\,\text{debye}$, which is in excellent agreement with the value presented in Ref. 39 for InAs QDs and comparable to that for InGaAs QDs.[40] To verify the single-photon nature of $X^0$ during the Rabi oscillation, an autocorrelation measurement was performed with $\pi$-pulse excitation, which corresponds to the full population inversion of the QD states. The result without background subtraction is shown in Fig. 6(b) as a semi-log plot. We observe the strong suppression of the coincidence at zero time delay with a $g^{(2)}(0)$ value as low as 0.008. This strong suppression of the multi-photon emission probability under $\pi$-pulse excitation indicates that the combination of the minimum pulse area excitation for the full population of the QD states and the quasi-resonant excitation is a superior method for realizing ideal single-photon sources.

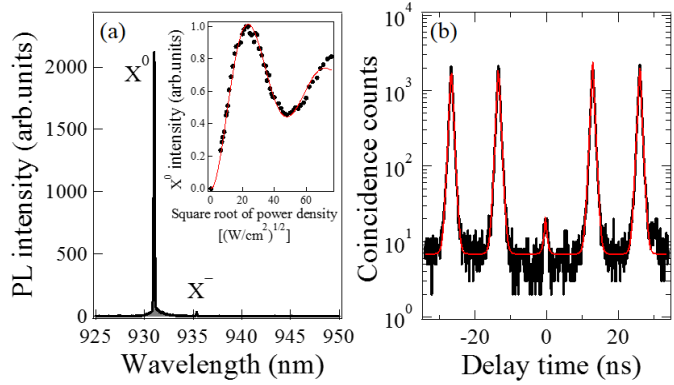

FIG. 6. (a) PL spectrum measured under the quasi-resonant excitation. The inset shows the Rabi oscillation of the $X^0$ emission intensity. (b) Semi-log plot of the measured second-order intensity correlation of the $X^0$ emission under the $\pi$-pulse excitation. The red curve is the theoretical fit to the data.[36]

## 7. Conclusion

We have combined the high source brightness and high indistinguishability of single photons generated from single InAs QDs embedded in Ag-coated GaAs nanocones. We have demonstrated a single-photon collection efficiency of 19%. We have also shown that the exciton states responsible for the single-photon emission can be switched from negatively charged to neutral excitons by changing from barrier excitation to first-excited-state quasi-resonant excitation. The ultrahigh purity of single photons was demonstrated with a $g^{(2)}(0)$ value as low as 0.008. The high indistinguishability of emitted photons was demonstrated with the help of Hong-Ou-Mandel-type two-photon interference. Although the measured two-photon interference visibility is limited by the time resolution of the instrument, an intrinsic visibility of 0.89 was derived. Furthermore, coherent population flopping of the exciton states was observed with the Rabi oscillations in the QD. To further improve both the source brightness and the photon indistinguishability, we could employ metallic nanocavities, in which the QD line can be coupled to the cavity mode.[41]

**ACKNOWLEDGEMENTS**

This work was partly supported by the Grand-in-Aid for Scientific Research (S), No. 24226007, Nanotechnology Platform, and Nano-macro materials, devices and systems alliance by the Ministry of Education, Culture, Sports, Science and Technology, and SCOPE (Strategic Information and Communications R&D Promotion Programme) from the Ministry of Internal Affairs and Communications, Japan.

**Figure captions**

FIG. 1. Schematic of the measurement setup.

FIG. 2. (a) PL spectrum measured from a single QD at 20K. The emission lines $X^0$ and $X^-$ correspond to the neutral exciton and the negatively charged exciton, respectively. (b) Time-resolved PL decay of the $X^-$ line. The solid curve is the convolution of single exponential decay with IRF. The IRF shown in the inset is fitted with Gaussian function. (c) HBT setup based on optical fibers for measuring second-order correlation function. An optical delay of 25 ns was inserted. TCSPC represents the time-correlated single-photon counting. (d) Normalized coincidence of the $X^-$ line as a function of the delay time. The solid curve is the fit to the experimental data.

FIG. 3. (a) Experimental setup of Michelson interferometer for measuring the coherence time. NPBS: non-polarizing beam splitter. The phase shifter was introduced to record the interference fringes at a fixed optical-path length difference set by the translation stage. (b) The visibility as a function of delay time for the $X^-$ emission. The inset shows an example of interference fringes at a position close to zero time delay.

FIG. 4. (a) Schematic of CW two-photon interference measurement. A fixed time delay between two optical paths was 5.8 ns. (b) and (c) are the normalized two-photon interference coincidences of the $X^-$ line under parallel and orthogonal polarizations, respectively. The inset in (b) is the zoom-in near the central dip. The horizontal dashed line corresponds to the $g^{(2)}(0) = 0.5$ level. The solid curves in (b) and (c) are calculated for parallel and orthogonal polarizations, respectively, with Eqs. (3) and (4) by taking the convolution with the IRF. The dashed curve in (b) is the calculation without the convolution.

FIG. 5. PLE spectra measured on $X^0$ and $X^-$ emissions. It should be noted that the PLE spectrum of the $X^-$ line is multiplied by 5.

FIG. 6. (a) PL spectrum measured under the quasi-resonant excitation. The inset shows the Rabi oscillation of the $X^0$ emission intensity. (b) Semi-log plot of the measured second-order intensity correlation of the $X^0$ emission under the $\pi$-pulse excitation. The red curve is the theoretical fit to the data.[36]